# What is digital about abstraction?

Bernhard Rieder, University of Amsterdam, Mediastudies Department

## Introduction

Abstraction is a central concept in computing in several ways. On one level, it is associated with "computational thinking" (Wing, 2006) and the "ability to perform abstract thinking" (Kramer, 2007), which connects to the question of which (cognitive) skills programmers need to master and what to teach in computer science education, calling attention to the epistemological content and character of the discipline. On another level, however, abstraction evokes *material* properties of digital computers that have significant consequences for how software is built and how the general capabilities of these machines are transformed into the myriad applications and services that have become fundamental parts of everyday life. Materiality, here, does not (necessarily) refer to hardware but points to the idea that computing, including software, is full of "stuff" (Dourish, 2016) that has particular characteristics and properties, a "substance" that enables and facilitates, but also orients and constrains what can be done. A programming language, for example, gives a *specific* material expression to the basic phenomenon of computing, to the point where Edsger Dijkstra, a key figure in the history of programming, argued that "the tools we are trying to use and the language or notation we are using to express or record our thoughts are the major factors determining what we can think or express at all" (1972: 864). Abstraction, enabled by digital representation and algorithmic control, is one of the central mechanisms for building and designing the material strata that define computing at any given point in time.

In this chapter, I focus on fleshing out this second understanding of abstraction and on connecting it with specific cultural and economic forms that have emerged in and around computing, including phenomena as contradictory as open-source software and the increasing "platformization" (Helmond, 2015) observable in various sectors of the industry. As computing continues to spread throughout society, the different kinds of layering and modularity we associate with abstraction affect an ever-growing number of domains and practices, reconfiguring how they are organized and how power relations emerge.

To make this argument, I first approach abstraction through a conceptual lens attentive to the historical trajectories that have shaped the specific forms it takes in computing practice today. I then connect abstraction with cultural consequences on the level of software-making, reflecting specifically on the meaning of layering and modularity. Finally, I discuss these ideas in relation to broader developments in the computing landscape, particularly the different instances of platformization we can observe in areas such as app development and artificial intelligence.

## The basics of abstraction

While abstraction is generally understood as a process of generalization, of deriving general rules or concepts from a broader space of cases or examples, it takes a specific meaning in computing, which ties it back to the fundamental features of digitality discussed in previous chapters. Digital computers represent both instructions and data as discrete symbols, the former defining algorithmic manipulations that operate on the latter. However, since instructions are also stored as symbols, they can be treated as data, and data can be used to generate instructions. We can thus imagine a program that reads a file containing sequences of symbols satisfying certain formal conditions and translates them into machine instructions. When it reads the word `print("hello world")`, for example, it generates the lines of machine code necessary to make these specific characters appear on the screen of the specific physical computer it runs on. This is how "high-level" programming languages work, and it is revealing that the earliest "translator" programs – which are now referred to as compilers or interpreters depending on how and when they translate – were called *autocoders*: they generate "actual" machine code automatically from some kind of "pseudo" code. This can only work when the latter correctly *abstracts* the former, that is, when the high-level programming language is able to map onto the logical structure of the underlying mechanisms, for example, a physical processor, a particular storage device, or a piece of network equipment. The higher level layers on top of the lower level and makes its functions available in generally more succinct and accessible form. Abstraction, understood in this sense, is a potent mechanism that draws on both discrete digital representation and algorithmic control to create all kinds of functional structures that programmers can draw on when making software. Understanding abstraction is not only essential to apprehending the specific materialities of computing but also to recognizing how it operates as an industrial field and, by extension, a cultural domain.

A central principle of abstraction is *information hiding*. In practice, this generally means that a simpler structure is built on top of a more complicated one, with the goal of making some functionality specified on the lower level available in a way that the technical details about how this functionality is achieved are not made visible. This makes it possible for programmers to write code for a computer architecture they know little about or to send and receive data over a network without having to understand the underlying communication protocols. What they need to know is the *interface* the lower level makes available to interact with it. Much of computing is indeed built on hierarchies of abstraction that can involve a surprising number of layers put on top of each other. The command `urllib.urlopen("http://myurl.com")`, for example, is used to read the contents of a web document into a script; it draws on the Urllib package, which is part of and built onto the Python programming language, which calls on network functions of the operating system, which implements a series of communication protocols and communicates with a device driver, which interfaces with hardware via a firmware interface, which communicates with a network chip somewhere on a motherboard or expansion card, which sends signals to another device, and so forth, and this is leaving out many steps in between. These strata have accumulated over decades of cumulative work in hardware and software, with the effect that reading a document from any server on the Internet is trivially easy for a Python programmer despite the mind-bending number of operations that happen in the background.

At the bottom of these sedimentations lies what Agre (1997) calls the "digital abstraction," which is realized by implementing the basic mechanisms of (today almost exclusively binary) computation in some physical medium. Charles Babbage's mechanical calculation engines were built with cogs and wheels, early digital computers used electromechanical relays, then vacuum tubes, and then transistors, and contemporary machines rely on integrated circuits, where billions of miniature elements are patterned and etched into silicon wafers. The abstraction happens when these material underpinnings are "forgotten," and the computer is treated as a device that can be understood in purely logical terms. From this point of view, all computing devices that are sufficiently "universal" – often called "Turing-complete" – are equivalent in the sense that they can calculate everything that is effectively calculable[1]. The fact that computers vary in terms of speed, memory size, reliability, energy use, and so forth, which has important consequences for their

---

[1] Computability theory is concerned with the question what can and cannot be effectively calculated; the point, here, is not to weigh in on this discussion but to emphasize that all computers that fulfill a number of basic requirements have, in theory, the same logical capacities.

*practical* capabilities, reminds us that "this abstraction from the material can never fully succeed" (Blanchette, 2011: 1042). More specifically, programmers indeed often find that all kinds of abstractions can be "leaky" (Spolsky, 2002) when the details of underlying implementation make themselves felt in one way or another.

Despite these caveats, "abstraction through information hiding is a primary factor in computer science progress and success" (Colburn and Shute, 2007), and the history of computing could very well be narrated as a history of cumulative abstraction. Stowing away the operational details of the underlying hardware is merely the ground floor of stratified infrastructures that are composed of layers upon layers of software. Assembly languages, which spread quickly from the late 1940s onwards, are abstractions built on the machine languages used to run early stored-program computers that are less cumbersome for human operators while remaining close to the underlying architectures. As computing spread and evolved further, it became clear that repetitive instruction patterns could be abstracted into compound commands similar to my Python examples, using terms from natural language to further facilitate programming practice. The high-level programming languages that came out of these efforts began to sever the close relationship with the hardware they ran on and could therefore be used across different computer architectures, with compilers translating programs into a variety of machine languages. As Dijkstra noted, "the purpose of abstracting is not to be vague, but to create a new semantic level in which one can be absolutely precise" (1972: 864), which means that higher level constructs are not ambiguous or polysemic like natural languages, but "black boxes" that compound a number of lower-level commands into a single statement. We may not know what something like `print("hello world")` in Python actually does in terms of the precise machine operations happening in our processor, but it will *reliably* display these two words on our screen when we run a script containing the line. The chains of translation abstraction is based on hold, at least most of the time.

Similar forms of abstraction can be found in other areas than programming proper, for example, in storage and networking. Keeping users from having to deal with physical locations on magnetic tape or rotating discs, the *file systems* appearing in the early 1960s started to provide abstractions – most importantly, the concept of the "file" itself – to facilitate data handling. Database systems later added another layer of abstraction on top of file systems. Edgar F. Codd famously introduced the "relational model" for database management, which seeks to fully separate the logical organization of data from its physical storage, with the declaration that "users of large data banks must be protected from having to know how the data is organized in the machine" (1970: 377). Similarly, *communication protocols* introduced concepts like the "packet" that hide away the realities of cables and electric signals (and many other things) in computer networks. The influential – but never fully implemented – Open Systems Interconnection (OSI) model goes as far as specifying seven layers of abstraction between physical transmission at the bottom and software application at the top to make the complicated task of reliable data transmission conceptually and practically manageable. In all of these cases, knowing how to use the higher-level constructs should be enough to work with them, and implementation details can be ignored – as long as they do not start to "leak," for example, when something goes wrong. As Latour writes,

> "When a machine runs efficiently, when a matter of fact is settled, one need focus only on its inputs and outputs and not on its internal complexity. Thus, paradoxically, the more science and technology succeed, the more opaque and obscure they become." (1999: 304)

Abstraction and black-boxing are, to a certain degree, also features of non-digital technology. It is not necessary to understand how a combustion engine works to successfully operate a motor vehicle. And cars are indeed made of thousands of increasingly standardized pieces built by subcontractors that operate more or less independently from the company assembling the final product. But mechanical systems simply do not have the flexibility and malleability afforded by digital systems and, in particular, software. Discrete representation and algorithmic control allow for forms of layering, stacking, and "building upon" that have no equivalent in physical space.

Operating systems such as Unix or Windows, which bundle thousands of abstractions together, have emerged as the most comprehensive "mediation layers" in computing. They not only hide and manage hardware details, but also keep programs running concurrently from bumping into each other, specify interaction principles through graphical user interfaces, and provide myriad system services that applications rely upon. While some caveats still apply, programmers working with languages such as Python today have little to no direct contact with hardware, at most, they have to address the idiosyncratic differences between the different operating systems on the market. The term "abstract machine" is often used in this context to address how a particular software environment (operating system plus programming language plus software packages plus …) comes to represent the computer as it appears for application programmers.

A central theme in the discussion around these abstraction processes has been the trade-off between the ease and flexibility gained by making complex functionality available through simple programming interfaces and the loss of performance and efficiency that can ensue (Agre, 1997). Since abstraction means that code is no longer fully tailored to the underlying hardware, optimal resource use can no longer be assured, and processing cycles are inevitably "wasted" on translation processes. The ability to write software faster and with greater ease comes at a price. In many ways, it is the steady progress in computing power, memory size, networking speed, and so forth, that has allowed for abstraction to proliferate. Dijkstra saw a fundamental relationship between hardware capacity and complexity, contending that "as the power of available machines grew by a factor of more than a thousand, society's ambition to apply these machines grew in proportion" (1972: 861) in the sense that applications that were simply unthinkable with one generation of computers would become feasible with the next. The different kinds of abstraction patterns used everywhere in computing today were the main vectors for realizing these ambitions. The practical consequences of this development went far beyond the initial intention to make programming less cumbersome and more easily transferrable from one computer architecture to another.

## Abstraction as Practice

In the 1960s – and often in direct relation to diagnoses of a "software crisis"[2] – abstraction came to be seen as an essential tool for dealing with the ever-increasing size and complexity of computer programs. Abstraction, then, was no longer merely a way of putting the particularities of hardware out of sight and making programming easier, but a *design principle* often associated with "divide and conquer" strategies that seek to split larger problems into smaller sub-problems. According to the influential *Structure and Interpretation of Computer Programs* (Abelson et al., 1996: xviii), "[w]e control complexity by building abstractions that hide details when appropriate," and this now concerns not only the makers of programming languages or operation systems but programmers working on any kind of application. The goal is to make programs *modular*, that is, to "divide the software into components such that each component can be designed independently of the others" (Ghezzi et al., 2002: 86). High-level programming languages increasingly propose constructs for supporting this process, in particular after Dijsktra's (1968) famously called the Go To statement – a close analog to the way branching is realized in physical processors – "harmful" and proposed "structured programming" as the right way forward. Subroutines were given center stage in the late 1950s in the FORTRAN II and ALGOL 58 languages, LISP was fully designed around functions, and the object-oriented programming paradigm that became popular in the 1970s introduced even more powerful ways to organize code into circumscribed units of functionality used to assemble larger programs. These constructs give programmers the means to create their own layers of abstraction, to build their own "languages" in the sense that they can create vocabularies of compound function that can be called upon just like the elements made available by programming languages or operating systems, much like the Urllib package mentioned above.

---

[2] In 1968, the NATO famously convenes a two-week conference in Garmisch-Partenkirchen to address the software crisis, which Buxton summarizes by pointing to the shared observation that "[i]n many sensitive areas, software is late, expensive, unreliable and does not work to specification" (1978: 23).

While avoidance of redundancy and, consequently, easier debugging play a role here, organizational challenges faced by a consolidating software industry, which produces programs too large for a single person to write and manage, were central (Ensmenger & Aspray, 2002). Modularity, understood as a form of abstraction, means that "each programmer should be able to work on a component with as little knowledge as possible about how other members of the team are building their components." (Ghezzi et al., 2002: 86). Information hiding thus became a principle in the organization of labor, and the separation between technical and social modes of structuring programming work began to blur. Indeed, if software *engineering* – "a systematic, disciplined, quantifiable approach to the development, operation, and maintenance of software" (IEEE, 1990) – emerges as the main response to the software *crisis*, the organizational or "managerial" dimension is more relevant than the technical one, even if the two necessarily intertwine.

Abstraction thus changes the character of computers as *means of production*, in particular, as means of production of other software. As both the technical and organizational dimensions of computing develop towards increasing differentiation, enabled by information hiding and design patterns built around modularity, the programming office becomes a different kind of work environment or "factory" and raises questions about the local and global division of labor and its social consequences.

Modularity indeed needs to be understood in broader terms than merely on the level of subroutines or programming libraries. As Agre states, "[c]omputer engineering uses abstraction to organize its work into a hierarchy of relatively independent levels, each with its own research community" (Agre 1997: 85), and this has led to larger forms of division of labor than what is happening within companies. Indeed, the separation between some people building hardware and others building software has been true for computers early on, but specialization has become much more fine-grained as the field has continued to diversify. Modules, large and small, are therefore not only a means of dividing up work but also of transmitting the deep knowledge captured in specialized components. In fact, the above-mentioned performance trade-offs coming with abstraction were not only compensated by increasingly powerful and "layer-aware" hardware but also by the expertise that is captured at the lower layers. For example, even if a programmer can write assembly code, it is doubtful that they can surpass the code quality of a modern compiler, not even counting the inevitable slowdown in their work progress. Programming languages and libraries may translate and abstract at a cost, but they are also constantly optimized by highly skilled specialists seeking to improve performance. The idea that any individual or team could easily match the "frozen" knowledge even small pieces of software draw upon is simply no longer viable. Computing has become a truly collective endeavor.

The specific shape collectivity takes is variable, however. On the one side, abstraction and modularity have allowed for the emergence of software "cathedrals" (Raymond, 1999), where large companies create large pieces of software and complex technical anatomies are mirrored by complex managerial structures. On the other side, the free software and open-source movements have been similarly reliant on the capacity to split up larger projects into many smaller ones. As Benkler argues, successful peer-production ventures "must be divisible into components, or modules, each of which can be produced independently of the production of the others" (2022: 378f) to allow for distributed production, but also to facilitate re-integration and quality control. Abstraction does not have a singular effect on (economic) organization but rather enables increases in the size and complexity of software by facilitating different forms of division of labor. Taylorist modes of production and exploitative forms of globalization can be seen as effects of abstraction, but so can participatory design, agile development, and the capacity for a single developer to produce truly impressive functional assemblages.

While the "proprietary" or "closed" vs. "free" or "open" software story has been told many times, abstraction has also facilitated the emergence of what economists call "ecosystems," that is, "group[s] of interacting firms that depend on each other's activities" (Jacobides et al., 2018: 2256). Modularity is again seen as central, here, because "it allows a set of distinct yet interdependent organizations to coordinate without full hierarchical fiat" (Jacobides et al., 2018: 2260) through well-defined patterns of interaction or interoperability between components. Instead of a binary opposition between open and closed, an ecosystems perspective detects various complex relationships between actors that can be characterized by different forms of inequality and domination. While certain dependencies exist in both cases, the relationship between game developers and leading game engines like Unreal Engine is different from the way app developers for smartphones are reliant on the Apple App Store or the Google Play Store. Through this lens, even the open/free software movement appears as a much more complicated environment than a set of egalitarian peer-production projects, as different kinds of companies have come to play leading roles in financing, contributing to, or controlling development. The Android operating system may be, in large parts, open source, but that does not mean that it is not heavily controlled by Google (Spreeuwenberg & Poell, 2012) and is itself a tool for controlling an entire ecosystem. Inquiring into the specific roles specific abstractions play in specific settings can allow us to remember that they "are created through the erasure of context, of specificity, of people, of labor, of experiments on paper, of company bottom lines, of origins, of failed alternatives" (Abbate & Dick, 2022: 439). Taking abstractions as entry points into the study of computing infrastructures is thus a valuable direction for critical scholarship.

## Abstraction as Platform

To question abstraction in terms of power is certainly not new. Friedrich Kittler, who lamented (1993: 209) that Microsoft switched from machine code to assembly language in their low-level documentation materials, saw abstraction mainly as a means to remove users' control over their machines and to re-introduce extractive principles such as ownership and copyright into the domain of software. For him, the accumulation of layers and associated control mechanisms represented a "postmodern Tower of Babel" (1993: 228) that did nothing but hide away the actual functioning of hardware, without providing any real utility. While Kittler's critique may seem overly idealistic or even elitist, it is indeed hard to deny that abstraction and power have become intimately entangled. In this last section, I want to discuss abstraction through the notions of "platform" and "platformization" to sharpen our view for contemporary forms this relationship has taken.

When looking at the history of the field often referred to as "platform studies," at least two main directions emerge. A more technical strand focused on "computer system[s] of any sort upon which further computing development can be done" (Bogost & Montfort, 2009: 1) strays close to the preceding discussion of abstraction in the sense that these systems, for example, gaming consoles, provide a platform for developers to build on. Here, hardware and software provide an abstract machine that is "universally" programmable but also offers certain capabilities, limitations, and idiosyncrasies that circumscribe the kinds of applications that can be built on it. This also connects to early discussions about changes in the way software is developed and deployed prompted by the internet: O'Reilly's (2005) original text on "Web 2.0" emphasizes the web as a development platform, and Bogost & Montfort (2009) point to the emergence of Web-APIs as ways for one system to draw on functionality or data made available by another. While we can still think of these interfaces as forms of abstraction that provide clear interaction patterns but hide what happens inside, we move away from a setup where a single computer sits below the various layers. A second strand present in platform studies emphasizes the relational aspects the term evokes, following either a more restricted economic definition, where platforms are basically synonymous with multi-sided markets (Rochet & Tirole, 2003), or a broader understanding, often adopted in fields like media studies, that can include different kinds of services as long as they "afford an opportunity to communicate, interact, or sell" (Gillespie, 2010: 351). In both cases, *intermediation* is the focus of attention and, in particular, the question of how intermediation is governed through business models, interfaces, algorithmic ordering, terms of service, and so forth. While programmability is central to the first understanding of platform, it is no longer a definitional requirement for the second.

When discussing abstraction, programmability is clearly essential, but the entanglement between technical infrastructure and economic organization is crucial for understanding the consequences of contemporary extensions to the concept. However, the platformization of computing is a long-standing development. If abstraction, in the sense discussed here, can always be understood as creating a platform for others to build on, operating systems have long been prime examples of the way technical and economic dimensions intertwine, as they set not only a technical stage but figure prominently in discussions of multi-sided markets that connect applications developers with audiences of users (Rochet & Tirole, 2003). Web-APIs have received much attention in recent years, but networked service provision has been a part of the internet from the beginning, even if rampant commercialization is indeed a newer development. Time-sharing, which allowed for the collective use of forbiddingly expensive resources in the mainframe era, is ultimately not that different from contemporary developments around cloud computing and virtualization.

In all of these areas, there have been many important developments that should not be minimized, but the most remarkable changes that we have seen in the last decade are not to be found in these individual trajectories but in their progressive combination and integration. iOS and Android are in many ways traditional operating systems and, at their core, just variants of Unix; but the tight coupling between hardware and software, the limited flexibility when it comes to choosing development environments and deployment pathways, the massive presence of integrated services, the many kinds of security measures, and the deep integration between the operating system and monetization possibilities for app developers are creating software platforms that are different from what we have come to know in the PC era. Abstraction, here, is pushed to new heights, as applications are confined to runtime environments that grant no direct access to the underlying hardware, and, more importantly, information hiding is extended beyond the purely technical. The app stores, in particular, have taken over most parts of payment and deployment, transforming these processes from direct interactions with end users into yet another layer addressable via some kind of API call. The details of how apps are hosted and how credit cards are charged are no longer in view, keeping with the push towards simplicity that has driven abstraction from the beginning.

Cloud computing platforms constitute another example of increased integration. While the basic principles are indeed similar to time-sharing, virtualization has allowed for forms of dynamic scaling that are clearly different. The capacity to move an operating system instance or a software container from one machine to another, possibly multiplying compute resources by orders of magnitude in the processes, has led to a "global restructuring in the composition, organization, and consumption of computing assets" (Narayan, 2022: 923), allowing companies to scale their businesses quickly in response to user demand, but also further cementing the outsized role Big Tech companies have come to play in computing today. However, in line with my previous argument, the most substantial development, here, is maybe not the fluid provision of colossal amounts of compute capability, but the (vertical) integration of hardware and software services. Since hardware is at least somewhat interchangeable and moving from one cloud provider to another would be too easy, the biggest players – Amazon Web Services, Microsoft Azure, and Google Cloud Platform – are investing heavily in software components and services that can be made available conveniently through programming interfaces, making it harder to switch providers. This can come in the form of custom frameworks for all kinds of computing tasks, for example, machine learning or data warehousing, as well as services similar to the payment layers popularized by app stores. In the end, the goal is to provide a development infrastructure – or abstract machine – that is filled to the brink with functionalities that developers and companies can draw on to make their work easier, increasing their dependency and making defection more costly.

In all of this, abstraction is caught up in power relations that keep evolving, but always operate through the ambiguity between technical and social forms and principles. Division of labor is particularly important. The initial split between hardware- and software-makers, the differentiation of designers and users of programming languages, the emergence of operating systems as comprehensive mediation layers built by teams of thousands, the rise of cloud computing, and the progressive integration of service layers, all of these – and other – developments imply moments of domination, or at least imbalance, in the sense that the "lower layers" I have talked about extensively shape (but not determine) what happens further up. As Giddens (1990) argues, having to constantly put trust into systems that we neither control nor understand is a "condition of modernity" and not exclusive to technology. The question, then, is not how to escape the fundamental interdependence that we find in computing and elsewhere, but how to arrange it in forms that promote fundamental rights and freedoms.

The main worry, today, should not be that software developers and end users depend on the work of other people or organizations, but that a small number of influential companies have come to define the "abstraction landscape" to a high degree. Certainly, companies like IBM and Microsoft have held overwhelmingly dominant positions in the past, but at a point in time when computing had simply not yet infiltrated the pores of society to the extent it has today. Google, in many ways, faces more competition than IBM in the 1960s, but the sheer range of computing fields it exerts influence in is staggering. From web search and web standards to mapping, mobile and cloud computing, and the Internet of Things, the company exercises control over important product spaces, popular means of production (programming languages, frameworks, etc.), technical standards, and cutting-edge research (Ahmed & Wahed, 2020). Together, Big Tech companies have a controlling stake almost everywhere in computing, and since computing is everywhere, this concerns a very large number of domains.

Abstraction – and this is where Kittler's work remains pertinent – creates opportunities for interdependence and, thus, control. The capacity to put layers on top of layers has consequences for digital computers as both technical and social objects. The notion of materiality this chapter started out with is central. The computer, as a material object, enables the "digital abstraction" Agre (1997) describes, implementing both discrete representation and algorithmic control, and allowing for the emergence of strata of hardware and software that deploy their own material "insistence" when they form abstract machines and development environments for others to build upon. These creations then play their own roles in the social domains they integrate, introducing their capabilities, forms, limitations, idiosyncrasies, and "character" into these settings. The economic consequences I have focused on are only one angle amongst others, as the whole gamut of human experience is affected by computing. The challenge, however, is not to reduce abstraction to a single "effect," but to trace its influence through different phenomena and settings. This is how the historical imagination can sensitize our thinking for the complexities and contradictions of the present.

# References

Abbate, Janet, and Stephanie Dick, eds. 2022. *Abstractions and Embodiments: New Histories of Computing and Society*. Studies in Computing and Culture. Baltimore: Johns Hopkins University Press.
Abelson, Harold, Gerald Jay Sussman, and Julie Sussman. 1996. *Structure and Interpretation of Computer Programs*. 2nd ed. Cambridge, Mass. : New York: MIT Press ; McGraw-Hill.

Agre, Philip. 1997. *Computation and Human Experience*. Cambridge: Cambridge University Press.

Ahmed, Nur, and Muntasir Wahed. 2020. "The De-Democratization of AI: Deep Learning and the Compute Divide in Artificial Intelligence Research." *arXiv:2010.15581 [Cs]*, October. http://arxiv.org/abs/2010.15581.

Benkler, Yochai. 2002. "Coase's Penguin, or, Linux and 'The Nature of the Firm.'" *The Yale Law Journal* 112 (3): 369. https://doi.org/10.2307/1562247.

Blanchette, Jean-François. 2011. "A Material History of Bits." *Journal of the American Society for Information Science and Technology* 62 (6): 1042–57. https://doi.org/10.1002/asi.21542.

Codd, E F. 1970. "A Relational Model of Data for Large Shared Data Banks" 13 (6): 11.

Colburn, Timothy, and Gary Shute. 2007. "Abstraction in Computer Science." *Minds and Machines* 17 (2): 169–84. https://doi.org/10.1007/s11023-007-9061-7.

Dijkstra, Edsger W. 1968. "Letters to the Editor: Go to Statement Considered Harmful." *Communications of the ACM* 11 (3): 147–48. https://doi.org/10.1145/362929.362947.

Dijkstra, Edsger W. 1972. "The Humble Programmer.Pdf." *Communications of the ACM* 15 (10): 859–66.

Dourish, Paul. 2017. *The Stuff of Bits: An Essay on the Materialities of Information*. Cambridge MA: The MIT Press. https://doi.org/10.7551/mitpress/10999.001.0001.

Ensmenger, Nathan, and William Aspray. 2002. "Software as Labor Process." In *History of Computing: Software Issues*, edited by Ulf Hashagen, Reinhard Keil-Slawik, and Arthur L. Norberg, 139–65. Berlin, Heidelberg: Springer Berlin Heidelberg. https://doi.org/10.1007/978-3-662-04954-9_12.

Ghezzi, Carlo, Mehdi Jazayeri, and Dino Mandrioli. 2002. *Fundamentals of Software Engineering*. 2nd ed.; Eastern economy ed. New Delhi: Prentice-Hall of India.

Giddens, Anthony. 1990. *The Consequences of Modernity*. Cambridge: Polity Press.

Gillespie, Tarleton. 2010. "The Politics of 'Platforms.'" *New Media & Society* 12 (3): 347–64. https://doi.org/10.1177/1461444809342738.

Helmond, Anne. 2015. "The Platformization of the Web: Making Web Data Platform Ready." *Social Media + Society* 1 (2): 205630511560308. https://doi.org/10.1177/2056305115603080.

Jacobides, Michael G., Carmelo Cennamo, and Annabelle Gawer. 2018. "Towards a Theory of Ecosystems." *Strategic Management Journal* 39 (8): 2255–76. https://doi.org/10.1002/smj.2904.

Kittler, Friedrich A. 1993. *Draculas Vermächtnis: technische Schriften*. 1. Aufl. Reclam-Bibliothek 1476. Leipzig: Reclam.

Latour, Bruno. 1999. *Pandora's Hope: Essays on the Reality of Science Studies*. Cambridge, Mass: Harvard University Press.

Narayan, Devika. 2022. "Platform Capitalism and Cloud Infrastructure: Theorizing a Hyper-Scalable Computing Regime." *Environment and Planning A: Economy and Space* 54 (5): 911–29. https://doi.org/10.1177/0308518X221094028.

Raymond, Eric. 1999. "The Cathedral and the Bazaar." *Knowledge, Technology & Policy* 12 (3): 23–49.

Rochet, Jean-Charles, and Jean Tirole. 2003. "Platform Competition in Two-Sided Markets." *Journal of the European Economic Association* 1 (4): 990–1029. https://doi.org/10.1162/154247603322493212.

Spreeuwenberg, Kimberley, and Thomas Poell. 2012. "Android and the Political Economy of the Mobile Internet: A Renewal of Open Source Critique." *First Monday* 17 (7). https://doi.org/10.5210/fm.v17i7.4050.

Wing, Jeannette M. 2006. "Computational Thinking." *Communications of the ACM* 49 (3): 33–35.